\documentclass[11pt,aps,pra,showpacs,showkeywords]{revtex4-1}
\usepackage{amsmath}
\usepackage{amssymb}
\usepackage{bm}
\usepackage{graphicx}
\usepackage{lineno}
\usepackage{color}
\begin{document}
\title{The Wigner-Witmer diatomic eigenfunction}

\author{
J. O. Hornkohl, A. C. Woods and C. G. Parigger\footnote{Corresponding author: Christian G Parigger, cparigge@tennessee.edu}}

\address{The Center for Laser Applications, \\ The University of Tennessee Space Institute, \\ Tullahoma, TN 37388 - 9700, \\ U.S.A.}





\begin{abstract}
Born and Oppenheimer reported an approximate separation of molecular eigenfunctions into electronic, vibrational, and rotational parts, but at the end of their paper showed that the two angles describing rotation of the nuclei in a diatomic molecule are exactly separable. A year later in a two-part work devoted strictly to diatomic molecules, Wigner and Witmer published (1) an exact diatomic eigenfunction and (2) the rules correlating the electronic state of a diatomic molecule to the orbital and spin momenta of the separated atoms. The second part of the Wigner-Witmer paper became famous for its correlation rules, but, oddly, the exact eigenfunction from which their rules were obtained received hardly any attention. Using three fundamental symmetries, we give a derivation of the Wigner-Witmer diatomic eigenfunction. Applications of our derivations are fundamental to predicting accurate diatomic molecular spectra that we compare with recorded spectra for diagnostic purposes, such as measurements of molecular spectra following generation of laser-induced plasma.
\end{abstract}
\maketitle

\section{Introduction}
In the introduction to their paper, Born \& Oppenheimer \cite{Born&Oppenheimer} allude to an exact separation of two rotational coordinates in the diatomic molecule. In their next section, which is applicable to polyatomic molecules, they introduce a coordinate system attached to the nuclei whose orientation is set by the Euler angles, and note that there are terms in the molecular Hamiltonian in which both electronic and nuclear coordinates appear thereby preventing the exact separation of the total eigenfunction into a product of electronic and nuclear eigenfunctions. In their final section, Born and Oppenheimer return to the diatomic molecule and give the details of the exact separation of two of the Euler angles, angles $\theta$ and $\phi$ that describe rotation of the two nuclei. The spherical harmonic $Y_{\ell \, m}(\theta, \phi)$ is the angular momentum part of Born-Oppenheimer diatomic eigenfunction. A year after, Wigner \& Witmer \cite{Wigner&Witmer} published a two-part article on diatomic theory in which they replace the spherical harmonic $Y_{\ell \, m}(\theta, \phi)$ with the Wigner $D$-function $D_{M \Omega}^{J^{\scriptstyle *}} (\phi, \theta, \chi)$. The Wiger-Witmer paper\cite{Wigner&Witmer} was the introduction of the rotation matrix elements into diatomic theory. After the Wigner-Witmer paper, the Wigner $D$-function mostly disappeared from diatomic literature for about four decades. Hirschfelder \& Wigner \cite{Hirschfelder&Wigner} used the $D$-function in their separation of 6 coordinates (3 for the total linear momentum, 3 for the total angular momentum) for $N$ particle systems, but do not explicitly mention the diatomic molecule. Again not specifically mentioning the diatomic molecule, Curtis  \& Hirschfelder \& Adler \cite{Curtis&2} repeat the separation of 6 coordinates for $N$ particle systems, and consider the three-body system in detail. Davydov \cite{Davydov} in his quantum mechanics textbook used the $D$-function in his discussion of diatomic theory. At about the same time, Rubin \cite{Rubin} employed it for his calculations of H\"onl-London factors. Pack \& Hirschfelder \cite{Pack&Hirschfelder} used the $D$-function to separate two angular rotational coordinates in the diatomic eigenfunction but failed to notice that their Eq.~(2.35) holds for all values of the third Euler angle, not just $\gamma=0$. Zare \cite{Zare1973} explicitly used the $D$-function in their case (a) basis function. Judd \cite{Judd} and Mizushima \cite{Mizushima}, in their treatments of diatomic theory, introduce the $D$-function and discuss its mathematical properties, but do not explicitly display it in their Hund's cases (a) and (b) basis functions. The Wigner $D$-function has since become a vital mathematical component of diatomic theory, as comprehensively collated by Varshalovich \& Moskalev \& Khersonskii \cite{Varsha}. However, the exact separation of $D_{M \Omega}^{J^{\scriptstyle *}} (\phi \, \theta \, \chi)$ in the diatomic eigenfunction where the nuclear coordinates $\phi$ and $\theta$ appear with the electronic coordinate $\chi$ has remained essentially forgotten for eight decades. We give a derivation of the Wigner-Witmer diatomic eigenfunction based upon three fundamental symmetries and the geometrical symmetry of a molecule possessing precisely two nuclei.

\section{Derivation of the Wigner-Witmer diatomic eigenfunction.} \label{WWequation}
Here we obtain the Wigner-Witmer eigenfunction by applying three symmetry principles to the eigenfunction of a free conservative system composed of $N$ electrons and precisely two nuclei. Energy is the generator of translations in time, the time translation (evolution) operator $U(t,t_0)$ is a continuous unitary operator, the total energy is a constant of the motion, and the dependence of the eigenfunction on the physical variable time $t$ is exactly separable if the time origin $t_0$ can be associated with some physical event. Linear momentum is the generator of translations in space, the spatial translation operator $\mathcal{T}(\mathbf{R},\mathbf{R}_0)$ is a continuous unitary operator, the total linear momentum is a constant of the motion, and the total linear momentum is exactly separable if the coordinates $\mathbf{R}_{\rm CM}$ of the center of mass can be introduced as physical variables of the system. Angular momentum is the generator of rotations, the rotation operator $\mathcal{R}(\alpha,\beta,\gamma)$ is a continuous unitary operator, the total angular momentum is a constant of the motion, but the total angular momentum $\mathbf{J}(\phi,\theta,\chi)$ is \emph{not}, in general, exactly separable because except for very simple systems one cannot find physical rotations $\phi$, $\theta$, and $\chi$ which duplicate the angles $\alpha$, $\beta$, and $\gamma$ of coordinate rotation. The diatomic molecule deserves a special place in the quantum theory of angular momentum because it is one of the most complicated systems for which the Euler angles $\alpha$, $\beta$, and $\gamma$ of coordinate rotation are also the angles of physical rotation describing the total angular momentum $\mathbf{J}$

Quantum mechanical descriptions of the diatomic molecule typically begin with the Hamiltonian, but minutia in the Hamiltonian tend to obscure the few fundamentals at play. For example, Brown \& Carrington \cite{Brown&Carrington} write a diatomic Hamiltonian, their Eq.~(2.297), containing 32 types of Hamiltonian terms. We begin our discussion of diatomic theory with the eigenfunction
\begin{equation}
\Psi_{nvJM}(\mathbf{R}_1, \mathbf{R}_2, \dots, \mathbf{R}_N, \mathbf{R}_{\rm a}, \mathbf{R}_b,t) \equiv \langle \mathbf{R}_1, \mathbf{R}_2, \dots, \mathbf{R}_N, \mathbf{R}_{\rm a}, \mathbf{R}_b,t \, |nvJM\rangle
\label{dm1}
\end{equation}

\noindent in which $\mathbf{R}_1, \mathbf{R}_2, \dots, \mathbf{R}_N$ are the spatial coordinates of the $N$ electrons and $\mathbf{R}_{\rm a}$ and $\mathbf{R}_b$ are those of the nuclei. The total angular momentum quantum numbers $J$ and $M$ refer to the true total. That is, in spectroscopic nomenclature they would be replaced by $F$ and $M_F$. The symbol $n$ represents all other required quantum numbers and continuous indices except the vibrational quantum number $v$.

The symmetries of translation in time and translation in space produce a separation of the time coordinate $t$ and the spatial coordinates $\mathbf{R}_{\rm CM}$ of the center of mass. A two-body reduction of the motion of the nuclei requires placement of the coordinate origin a the center of mass of the nuclei, and then replaces $\mathbf{R}_{\rm a}$ and $\mathbf{R}_{\rm b}$ with the internuclear vector $\mathbf{r}$. Of the $3N+7$ dynamical variables in the total eigenfunction (\ref{dm1}), $3N+3$ remain in the internal eigenfunction $\langle \mathbf{r}_1, \mathbf{r}_2, \dots, \mathbf{r}_N,\mathbf{r} \, |nvJM\rangle$. The axes of the translated coordinates $\mathbf{r}_1, \mathbf{r}_2, \dots, \mathbf{r}_N,\mathbf{r}$ are parallel to those of the original laboratory coordinates $\mathbf{R}_1, \mathbf{R}_2, \dots, \mathbf{R}_N, \mathbf{R}_{\rm a}, \mathbf{R}_{\rm b}$. We now address how rotational symmetry influences the internal eigenfunction.

Operation of the rotation operator $\mathcal{R}(\alpha,\beta,\gamma)$ on the internal eigenfunction yields
\begin{equation}
\langle \mathbf{r}_1, \mathbf{r}_2, \dots, \mathbf{r}_N,\mathbf{r} \, | \mathcal{R}(\alpha,\beta,\gamma) |nvJM\rangle = \langle \mathbf{r}'_1, \mathbf{r}'_2, \dots, \mathbf{r}'_N,\mathbf{r}' \, | nvJM\rangle
\end{equation}

\noindent where primes denote rotated coordinates given by
\begin{equation}
\mathcal{D}(\alpha, \beta, \gamma)
= \left[ \begin{matrix} \cos\alpha \cos\beta \cos\gamma - \sin\alpha \sin\gamma & \sin\alpha \cos\beta \cos\gamma + \cos\alpha \sin\gamma &  -\sin\beta \cos\gamma \\
-\cos\alpha \cos\beta \sin\gamma - \sin\alpha \cos\gamma & -\sin\alpha \cos\beta \sin\gamma + \cos\alpha \cos\gamma & \sin\beta \sin\gamma \\
\cos\alpha \sin\beta & \sin\alpha \sin\beta &  \cos\beta
\end{matrix} \right], \label{Drot}
\end{equation}
\begin{equation}
\left[ \begin{matrix}
x' \\
y' \\
z' \end{matrix} \right]
= \mathcal{D}(\alpha, \beta, \gamma)
\left[ \begin{matrix}
x \\
y \\
z \end{matrix} \right].
\label{r'}
\end{equation}

\noindent The effect of $\mathcal{R}(\alpha,\beta,\gamma)$ on the eigenfunction can be rewritten as
\begin{align}
\langle \mathbf{r}_1, \mathbf{r}_2, \dots, \mathbf{r}_N,\mathbf{r} \, | nvJM\rangle &= \langle \mathbf{r}'_1, \mathbf{r}'_2, \dots, \mathbf{r}'_N,\mathbf{r}' \, | \mathcal{R}^{\dagger}(\alpha,\beta,\gamma) |nvJM\rangle \\
&= \sum_{\Omega=-J}^J \langle \mathbf{r}'_1, \mathbf{r}'_2, \dots, \mathbf{r}'_N,\mathbf{r}' \, |nvJ\Omega\rangle \, \langle J\Omega \, |  \mathcal{R}^{\dagger}(\alpha,\beta,\gamma) \, | JM\rangle \\
&= \sum_{\Omega=-J}^J \langle \mathbf{r}'_1, \mathbf{r}'_2, \dots, \mathbf{r}'_N,\mathbf{r}' \, |nvJ\Omega\rangle \, D_{M \Omega}^{J^{\scriptstyle *}} (\alpha, \beta, \gamma).
\end{align}

\noindent When the spherical coordinates of the internuclear vector $\mathbf{r} = \mathbf{r}(r,\theta,\phi)$ are introduced, the equation becomes
\begin{equation}
\langle \mathbf{r}_1, \mathbf{r}_2, \dots, \mathbf{r}_N,r,\theta,\phi \, | nvJM\rangle \\
= \sum_{\Omega=-J}^J \langle \mathbf{r}'_1, \mathbf{r}'_2, \dots, \mathbf{r}'_N,r,\theta',\phi' \, |nvJ\Omega\rangle \, D_{M \Omega}^{J^{\scriptstyle *}} (\alpha, \beta, \gamma).
\end{equation}

\noindent The internuclear distance $r$ is unprimed on the right because it is a scalar. Because physical rotation $\phi$ and coordinate rotation $\alpha$ are both counterclockwise rotations about the $z$ axis, the physical angles $\phi'$ is given by
\begin{equation}
\phi' = \phi - \alpha.
\end{equation}

\noindent Similarly, physical rotation $\theta$ and coordinate rotation $\beta$ are counterclockwise rotations about the first intermediate $y$ axis of the total coordinate rotation.
\begin{equation}
\theta' = \theta - \beta.
\end{equation}

\noindent Rotational symmetry gives us the option to view the molecule at any orientation we choose, and we choose $\alpha=\phi$ and $\beta=\theta$.
\begin{equation}
\langle \mathbf{r}_1, \mathbf{r}_2, \dots, \mathbf{r}_N,r,\theta,\phi \, | nvJM\rangle = \sum_{\Omega=-J}^J \langle \mathbf{r}'_1, \mathbf{r}'_2, \dots, \mathbf{r}'_N,r \, |nvJ\Omega\rangle \, D_{M \Omega}^{J^{\scriptstyle *}} (\phi, \theta, \gamma).
\end{equation}

\noindent The rotated coordinates of the one of the electrons, we arbitrarily select the electron labeled 1, are expressed in cylindrical coordinates $\rho'_1$, $\zeta'_1$, and $\chi'_1$.
\begin{equation}
\langle \mathbf{r}_1, \mathbf{r}_2, \dots, \mathbf{r}_N,r,\theta,\phi \, | nvJM\rangle = \sum_{\Omega=-J}^J \langle \rho'_1, \zeta'_1, \chi'_1, \mathbf{r}'_2, \dots, \mathbf{r}'_N,r \, |nvJ\Omega\rangle \, D_{M \Omega}^{J^{\scriptstyle *}} (\phi, \theta, \gamma).
\end{equation}

\noindent The chosen electron is distance $\rho'_1$ from the internuclear vector and signed distance $\zeta'_1$ above or below the plane perpendicular to the internuclear vector and passing through the coordinate origin. The angle $\chi'$ describes rotation of this electron about the internuclear distance. Like the internuclear distance $r$, primes on $\rho'_1$ and $\zeta'_1$ are unnecessary because they are scalars whose values are unchanged by coordinate rotation. Because $\chi'_1$ and $\gamma$ are rotations about the same axis, coordinate rotation changes, of course, the value of $\chi'_1$, but this also means that this angle has a value $\chi_1$ in laboratory coordinates. The coordinate rotation angle $\gamma$ is chosen to make $\chi'_1$ zero,
\begin{equation}
\chi'_1 = \chi_1 - \gamma .
\end{equation}

\noindent The equation for the eigenfunction now reads
\begin{equation}
\langle \rho, \zeta, \chi, \mathbf{r}_2, \dots, \mathbf{r}_N, r, \theta, \phi \, | nvJM\rangle = \sum_{\Omega=-J}^J \langle \rho, \zeta, \mathbf{r}'_2, \dots, \mathbf{r}'_N,r \, |nv\rangle \, D_{M \Omega}^{J^{\scriptstyle *}} (\phi, \theta, \chi)
\label{WWeq}
\end{equation}

\noindent after the subscripts on $\rho_1$, $\zeta_1$, and $\chi_1$ have been dropped. This result is the Wigner-Witmer diatomic eigenfunction. The total diatomic eigenfunction is given as the sum of $2J+1$ products of electronic-vibrational basis functions $\langle \rho, \zeta, \mathbf{r}'_2, \dots, \mathbf{r}'_N,r \, |nv\rangle$ and total angular momentum basis functions $D_{M \Omega}^{J^{\scriptstyle *}} (\phi, \theta, \chi)$ or Wigner D-functions. The Born-Oppenheimer approximation separates the electronic-vibrational basis into the product of electronic and vibrational basis functions, but the separation of $D_{M \Omega}^{J^{\scriptstyle *}} (\phi, \theta, \chi)$ from the electronic-vibrational basis is exact. Many individual orbital and spin momenta are contained in the electronic-vibrational basis, but $D_{M \Omega}^{J^{\scriptstyle *}} (\phi, \theta, \chi)$ is the total angular momentum basis.

It is noteworthy that eigenfunctions for a rotational state of the diatomic molecule usually contain parity, in other words, parity symmetrization is customary. Inclusion of the discrete parity symmetry is accomplished after the construction of the Wigner-Witmer diatomic eigenfunction. The parity operator can be constructed from proper and improper rotations that have a determinant of the transformation matrix of +1 and -1, respectively. Subsequently, Equation 2.14 for the eigenfunction can be split for specific values of $\Omega$ followed by parity symmetrization. The approach of including parity for specific values of $\Omega$ has been utilized in the literature in order to reduce the size of the Hamiltonian matrix prior to finding eigenvalues by diagonalization. In our work \cite{JALSpaper}, the parity operation is considered after establishment of the eigenfunction in terms of $J$ and $M$ as sum over $\Omega$ in Eq. 2.14.

Clearly, as electronic states for the diatomic molecule are considered, parity is paramount for building these states utilizing the Wigner-Witmer correlation rules \cite{Bellary}. Yet in this work we focus on the use of the Wigner-Witmer eigenfunction for computation of spectra, rather than molecular structure predictions in non-Born-Oppenheimer calculations for molecules \cite{Adamowicz1,Adamowicz2}.

\section{Discussion and conclusion}
The mixing of the electronic coordinate $\chi$ with the two nuclear coordinates $\phi$ and $\theta$ in $D_{M \Omega}^{J^{\scriptstyle *}} (\phi, \theta, \chi)$ is an obvious departure from current expositions of diatomic theory. However, the exact separation of the total angular momentum basis from the electronic-vibrational basis proves useful. For example, writing the parity operator $\mathcal{P}$ as the product of a proper rotation $\mathcal{P}_{\alpha \beta \gamma}$ and an improper rotation $P_{\Sigma}$,
\begin{equation}
\mathcal{P} = \mathcal{P}_{\alpha \beta \gamma} \, P_{\Sigma},
\end{equation}

\noindent one obtains a simple equation for the parity of diatomic states. The eigenvalue of $\mathcal{P}$ is, of course, $\pm 1$, and the product of eigenvalues $p_{\Sigma} \, p_{\alpha \beta \gamma}$,
\begin{equation}
p = p_{\Sigma} \, (-)^{J+2M},
\label{diatomicParity1}
\end{equation}

\noindent is always $\pm 1$ as required. Sign changes due to parity can show different effects on the Wigner D-function.
Depending on the specific rotation group, the effect can be expressed in terms of $J$ and $M$ or in terms of $J$ and $\Omega$. Table \ref{C2}
summarizes the sign changes due to parity. These results are consistent with the ones presented by Varshalovich \& Moskalev \& Khersonskii \cite{Varsha}.
\begin{table}[!h]
\begin{center}
\caption{The sign changes on the components $x'$, $y'$ and $z'$ of a coordinate vector ${\bf r}'(x',y',z')$ produced by three different discrete Euler angle transformations, and the effect of these Euler angle transformations on $D_{M \Omega}^{J^{\scriptstyle *}} (\alpha, \beta, \gamma)$.}
{\begin{tabular}{llll}
\hline 
Transformation group & Euler angles & Coordinates &  Effect on $D_{M \Omega}^{J^{\scriptstyle *}} (\alpha, \beta, \gamma)$ \\
\hline 
$C_2(x')$ & $\alpha \rightarrow \pi + \alpha$ & $x' \rightarrow x'$ & \\
& $\beta \rightarrow \pi - \beta$ & $y' \rightarrow -y'$ & $(-)^{J+2M} \, D_{M,-\Omega}^{J^{\scriptstyle *}} (\alpha, \beta, \gamma)$ \\
&  $\gamma \rightarrow -\gamma$ &  $z' \rightarrow -z'$ & \\


$C_2(y')$  & $\alpha \rightarrow \pi + \alpha$ & $x' \rightarrow -x'$ & \\
& $\beta \rightarrow \pi - \beta$ & $y' \rightarrow y'$ & $(-)^{J-\Omega} \, D_{M,-\Omega}^{J^{\scriptstyle *}} (\alpha, \beta, \gamma)$  \\
& $\gamma \rightarrow \pi - \gamma$ & $z' \rightarrow -z'$ & \\


$C_2(z')$ &
$\alpha \rightarrow \alpha$ & $x' \rightarrow -x'$ & \\
& $\beta \rightarrow \beta$ & $y' \rightarrow -y'$ & $(-)^{-\Omega} \, D_{M,-\Omega}^{J^{\scriptstyle *}} (\alpha, \beta, \gamma) $ \\
& $\gamma \rightarrow \pi + \gamma$ & $z' \rightarrow z'$ & \\
\hline 
\end{tabular}}
\label{C2}
\end{center}
\end{table}

\noindent Note that for half-integer $J$ in Equation (3.2), the individual eigenvalues $p_{\Sigma}$ and $p_{\alpha \beta \gamma}$ are purely imaginary. A widely accepted convention allows one to treat the parity eigenvalues  $p_{\Sigma}$ and $p_{\alpha \beta \gamma}$ as real for both integer and half-integer $J$ \cite{Brown_ef1975}. If one agrees to always subtract $1/2$ from half-integer values of $J$, then the diatomic parity can be written as
\begin{subequations}
\begin{align}
p &= + p_{\Sigma} \, (-)^J \ \ \ \ \ \ \ \ \qquad J\text{ integer} \\
&= - p_{\Sigma} \, (-)^{J-1/2} \qquad J\text{ half-integer}
\end{align}
\label{diatomicParity2}
\end{subequations}

\noindent in which $p_{\Sigma} = \pm1$ is always real.

The Wigner-Witmer diatomic eigenfunction simplifies the process in which one infers molecular parameters such as the rotational parameter $B_v$ and the spin-orbit parameter $A_v$ from experimentally measured line positions. Application of our detailed Wigner Witmer eigenfunctions include analyses of low- and high-temperature spectra of diatomic carbon spectra \cite{JObook}, or as another example, development of line strengths for specific transitions of the aluminium monoxide (AlO) diatomic molecule \cite{SAA}.
The parameters we use are electronic-vibrational matrix elements. In turn, the Born-Oppenheimer approximation separates these into the products of electronic matrix elements times vibrational matrix elements, and introduces a large number of  differential equations which couple the many Born-Oppenheimer vibrational states thereby producing a large Hamiltonian matrix. Van Vleck transformations reduce the dimension of the Hamiltonian to yield an effective Hamiltonian. If in the fitting process, one deals with electronic-vibrational matrix elements such as $B_v$ and $A_v$ instead of breaking them into the products of electronic matrix elements and vibrational matrix elements, Van Vleck transformations are no longer required.

Conversely, the Wigner-Witmer eigenfunction does not reveal how the total angular momentum is built from its components. One must use an angular momentum coupling model which has a complete basis. For example, the Hund's case {\it a} basis appropriate to the Wigner-Witmer eigenfunction is
\begin{equation}
|a\rangle = |nJM\Omega \Lambda S \Sigma \rangle
=\sqrt{\frac{2J+1}{8 \pi^2}} \, \langle \rho, \zeta, \mathbf{r}'_2, \dots, \mathbf{r}'_N, r \, | nv \rangle \, D_{M \Omega}^{J^{\scriptstyle *}} (\phi, \theta, \chi) \, |S \Sigma \rangle ,
\end{equation}

\noindent where $\Omega = \Lambda + \Sigma$. Written in our notation, current practice replaces the above with

\begin{equation}
|a\rangle =  
\sum_{v_{BO}=0} \langle \rho, \zeta, \mathbf{r}'_2, \dots, \mathbf{r}'_N; r \, | n v_{BO} JM \Omega \Lambda \rangle \, \langle v_{BO} \, | v \rangle
D_{M \Omega}^{J^{\scriptstyle *}} (\phi, \theta, 0) \, |S \Sigma \rangle.
\label{BO}
\end{equation}

\noindent The total angular momentum is not exactly separated in this equation and there is the sum over a large number of Born-Oppenheimer vibrational states $|v_{BO}\rangle$. It is often said that the Born-Oppenheimer approximation separates the diatomic eigenfunction into electronic, vibrational, and rotational states, but this does not hold with spectroscopic accuracy. In the modeling of upper and lower Hamiltonians whose term differences accurately agree with measured line positions, one must deal with the large set of coupled differential equations that result when Eq.~(\ref{BO}) is inserted in the Schr\"odinger equation. Analytical techniques that employ Van Vleck transformations and parity symmetrization have been developed to yield much smaller effective Hamiltonians.

Except for the simplest of diatomic molecules, \emph{ab initio} computations are usually not as accurate as results obtained in experimental spectroscopy. For  an accurate prediction of a diatomic spectrum, one must have values for the molecular parameters such as $B_v$, $A_v$, $\lambda_v$, $\gamma_v$, $\dots,$  and their centrifugally stretched forms. Computer programs have been developed which find the molecular parameters by fitting upper and lower term differences from model Hamiltonians to accurately measured line positions. Such a program begins with trial values for the molecular parameters, computes theoretical line positions as eigenvalue differences between upper and lower Hamiltonians, computes corrections to the trial values of the molecular parameters from the differences between the computed and measured line positions, and iterates until corrections to the parameters become negligibly small. There are now many examples of this algorithm for which the errors in the computed line positions, \emph{i.e.,} computed vacuum wavenumbers, are not significantly larger then the estimated experimental accuracy. With two exceptions, replacement of the Born-Oppenheimer approximation with the Wigner-Witmer eigenfunction does not significantly alter the flow charts for these programs. First, the matrix elements of the effective Hamiltonian are replaced with Hund's case {\it a} matrix elements unmodified by Van Vleck transformations and parity symmetrization. Second, the manual enforcement of selection rules is replaced by computation of the H\"onl-London factors \cite{HLfactors}. The exact separation of the total angular momentum in the Wigner-Witmer eigenfunction provides simple, accurate computation of the H\"onl-London factors, and there is but a single diatomic selection rule: Transitions for which the H\"onl-London factor is non-vanishing are allowed. Thus, use of the Wigner-Witmer diatomic eigenfunction represents a significant departure from current practices in diatomic theory.


\end{document}